\documentclass[11pt,twoside]{article}

\usepackage{asp2004}
\usepackage{epsf}
\usepackage{lscape}
\usepackage{graphicx}

\begin{document}
\title{Pulsar Studies of Tiny-Scale Structure in the Neutral ISM}
\author{J. M. Weisberg}  
\affil{Department of Physics and Astronomy, Carleton College,
Northfield, MN 55057} 
\author{S. Stanimirovi\'{c} }  
\affil{Radio Astronomy Lab, UC Berkeley, 601 Campbell Hall,
Berkeley, CA 94720}

\begin{abstract} 

We describe the use of pulsars to study small-scale neutral structure in the interstellar medium
(ISM).  
Because pulsars are high velocity objects, the pulsar-Earth  line of sight sweeps rapidly across  
the ISM.  Multiepoch measurements of pulsar  interstellar spectral line spectra therefore probe 
ISM structures on AU scales. We review pulsar measurements of small scale structure in HI and 
OH and compare these results with those obtained through other techniques.
\end{abstract}



\section{Introduction}

The discovery of tiny-scale atomic structure (TSAS; \citet{h97}) in the neutral interstellar medium 
(ISM) was made in the 
VLBI neutral hydrogen (HI) absorption experiment of \citet{d76}.\footnote{See Brogan et al, 
this volume, for a review of 
interferometric studies of tiny-scale structure.}  Meanwhile, a complementary technique for
probing tiny-scale structure, using multiepoch pulsar observations, began  in the mid-1980s.
This technique takes advantage of pulsars' compact nature and their high velocities to probe the ISM 
on tiny scales by searching for temporal variations in spectral lines produced by the intervening
medium.  Until recently, these studies were confined to neutral hydrogen 21 cm absorption 
 lines, but in the
last few years the technique has been extended to investigations of OH in both absorption and
stimulated emission.

\section{The Technique}

Pulsars subtend very small angles on the sky, even after their intrinsic size is broadened by
interstellar scattering.  They tend to be high-velocity objects, with transverse speeds of 
hundreds of km s$^{-1}$.  Consequently the pulsar-Earth column  probed by a pulsar signal
is  both needle thin and is dragged  in short order across solar 
system size scales in the ISM. For example,  a 100 km s$^{-1}$ transverse pulsar velocity 
sweeps the line of sight across 20 AU in one year.  Multiepoch pulsar ISM measurements 
 at reasonable intervals then probe ISM structure on these scales.

In order to study the medium along the line of sight, one must measure the
 {\it{pulsar ISM spectrum}}\footnote{Historically the relevant spectrum was called the ``pulsar
 {\it{absorption}}'' spectrum, but recent observations of pulsar-stimulated OH {\it{emission}} require
 that the term ``absorption'' be replaced when describing this technique generically.} at and
near the frequency of a spectral line such as $\lambda\sim21$ cm HI or $\lambda\sim18$ cm OH.
The pulsar ISM spectrum, which is the spectrum of the pulsar {\it{alone}} (as modified by the
intervening ISM) is not a direct observable since the telescope samples a complicated
combination of signals throughout its beam.  Some form of beam switching or spatial filtering
is employed to separate the ISM spectrum of conventional steady sources.  However, 
because of the pulsed nature of pulsar 
emission, it is straightforward in principle to derive the pulsar ISM spectrum via {\it{temporal}} 
switching;  i.e., by differencing measurements of the spectrum when the pulsar 
is {\it{on}}, $I_{PSR-On}(\nu)$; and when it is   {\it{off}}; $I_{PSR-Off}(\nu)$:
\begin{equation}
I_{PSR-ISM}(\nu) =  I_{PSR-On}(\nu) - I_{PSR-Off}(\nu).
\end{equation}
Ultimately we desire the fractional absorption or fractional 
stimulated emission, $I_{PSR-ISM}(\nu)/ I_{PSR,o}$ where $I_{PSR,o}$ is the
unabsorbed pulsar intensity determined from off-line spectral channels, or equivalently
the (positive or negative) optical depth $\tau(\nu)$ of the gas along the pulsar-Earth line of sight, 
defined by 
\begin{equation}
\tau(\nu) =  -\ln(I_{PSR-ISM}(\nu)/ I_{PSR,o}).
\end{equation}
Early efforts utilized one-bit spectrometers whose output was gated synchronously with the pulsar
period into one of two buffers (pulsar-{\it{on}} or pulsar-{\it{off}}).  Calibration of the spectra was 
difficult because of  limitations inherent in one-bit  sampling of the wildly varying pulsar signal 
\citep{w80}.
Some of these limitations have been lessened through the years with the introduction of multibit
spectrometers whose output can be directed into multiple pulsar phase bins rather than only
two. The resulting multibit cube of data (intensity as a function of frequency and pulsar phase)
can then be optimally processed to yield the pulsar-{\it{on}}, pulsar-{\it{off}}, and pulsar ISM 
spectra.  Nevertheless, careful calibration
remains crucial since the desired pulsar ISM spectrum represents the (usually small) difference 
between two very similar spectra.

\section{Neutral Hydrogen Measurements and Analyses}

As hydrogen is the most pervasive constituent of the ISM, it is natural to study fluctuations in HI.  This
section examines the pulsar and interferometric HI experiments and analyses.

\subsection{History of Pulsar Tiny-Scale HI Structure Experiments}

Early efforts with the pulsar HI ISM technique focused on measuring pulsar distances kinematically.
This procedure continues to have great success, and provides primary calibration 
(e.g., \citet{fw90}; \citet{w96}) for models of the galactic density distribution  \citep{tc93,cl02}. 
By the late 1980s,
sufficiently accurate repeated measurements began to be made, and it was noticed that the 
pulsar ISM spectra changed over time in some cases, suggesting changes in the tiny-scale
structure of intevening gas.  For example, \citet{c88} found that 
the HI absorption spectrum of PSR B1821+05 changed significantly between ~1981 and 1988,
with the appearance at the latter epoch of a previously unobserved feature with $\tau\sim2$ and
$\Delta v \sim 1$ km s$^{-1}$.
 \citet{d92} showed that between $\sim1976$ and 1981, HI absorption toward  B1154-52 did 
{\it{not}} change significantly; while toward B1557-50, a variation with $\Delta\tau\sim1$ was
interpreted as a cloud of size in the 1000 AU range.

These early pulsar HI results inspired Frail, Weisberg, Cordes,  
\& Mathers (1994)  to mount a dedicated multiepoch pulsar HI absorption experiment at Arecibo.
Six pulsars were observed at three epochs, with time baselines ranging from 0.7  - 1.7 yr.  These
authors reported the presence of {\it{pervasive}} variations with $\Delta\tau\sim0.03-0.7$, 
and associated HI column densities of $10^{19} - 5 \times 10^{20}$ cm$^{-2}$; on scales of 5-100 
AU.  They indicated that 10-15\% of cold HI is in the tiny structures, and additionally detected
a correlation between absorption equivalent width variations and equivalent width itself.
These results appeared to buttress the earlier VLBI findings of small scale structure, and
they provided a strong impetus for further experimental and theoretical work.  

The theoretical work focused on explaining the very existence of TSAS. The 
presence of such structures  is
enigmatic, especially because they appear to be significantly overpressured
with respect to the ambient ISM.  \citet{h97} emphasized the seriousness of the overpressure
problem, and suggested that nonspherical cloud geometries such as filaments and sheets oriented
with their long axes along the line of sight could ameliorate it.  \citet{d00}  argued that the observed
changes do not result from discrete structures, but rather are merely the variations expected
from a reasonable
power-law spectrum of opacity fluctuations.  \citet{g01} showed that velocity gradients in the HI, 
combined with interstellar scattering of the pulsar signals, could explain the variations.

\subsection{Recent Pulsar HI TSAS Measurements}

The recent era of  pulsar TSAS experiments began with the Parkes observations
of  \citet{j03}.  Surprisingly, these
investigators found no significant optical depth variations in their three-epoch, 2.5-yr 
observations of three pulsars, and were able to place an upper limit on column density variations 
toward PSR B1641-45 of $10^{19}$ cm$^{-2}$, significantly below the 
Frail, Weisberg, Cordes, \& Mathers (1994)  detections.  They
showed that the earlier experiment did not fully account for the large increases in noise in  
absorption spectra at the line frequency, so that some of the apparently significant 
variations actually were not.  While no significant variations were seen by \citet{j03} during the
2.5 years of their TSAS experiment, they detected variations in the spectrum of PSR B1557-50 when
compared with measurements made five years earlier.  This is the same pulsar whose spectrum was noted earlier by \citet{d92} to vary on similar timescales in the late 1970s.  In combining the results from
four measurements over twenty-five years, \citet{j03} concluded that the cloud causing the variations
is $\sim 1000$ AU in size, with a density of $\sim 10^4$ cm$^{-3}$.

\citet{mET05} performed a particularly exhaustive TSAS study on PSR B0329+54, which is very
bright and almost circumpolar at the Green Bank Telescope. The investigators observed the pulsar
continuously for up to 20 h in eighteen observing sessions over a period of  1.3 yr.   They
detected no HI  opacity variations ($\Delta\tau < 0.026$ in most cases) for pulsar transverse
offsets ranging from 0.005 - 25 AU.  

\subsubsection{The New Arecibo Pulsar HI TSAS Experiment}

Recently,  we and our colleagues instigated a new set of Arecibo multiepoch pulsar HI TSAS studies.  
Advances in digital technology since the earlier observations of \citet{f94} allowed the use of a 
multibit spectrometer \citep{j01} whose calibration should minimize systematic errors; and receiver 
improvements decreased the noise.
Preliminary results were published by \citet{s03} and a final report is nearing completion \citep{s07}.  
We  observed the same six pulsars as \citet{f94} at four epochs, yielding time baselines between 0.2 
and 1.3 yr, with pulsars moving between 1 and 200 AU between sessions.  The line of sight
parameters for the six pulsars are given in Table 1. 

\begin{figure*}[!ht]
\includegraphics[scale=0.6,trim=-1in 0 0 0 ]{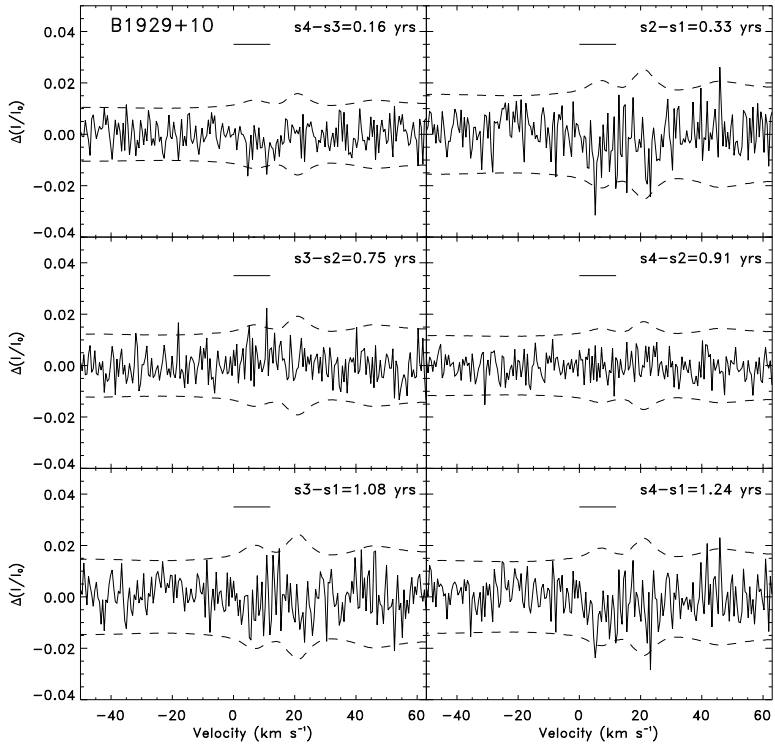}
\caption{Differences between pairs of HI absorption spectra for B1929+10. Data were gathered 
at Arecibo by \citet{s07}. The horizontal bar delineates the velocity range where significant 
absorption is present in 
single-epoch spectra. Dashed lines show $\pm2\sigma$ error envelopes. Note that only a few features
extend outside the envelopes.}
\end{figure*}

\begin{table}[!h]
\caption{Line of sight parameters in the Arecibo pulsar HI TSAS study}
\smallskip
\begin{center}
{\small
\begin{tabular}{cccccc}
\tableline
\noalign{\smallskip}
PSR J & PSR B &      $l$    & $b$    & $d$   & $v_{transverse}$ \\
            &               &   (deg)  &  (deg) & (kpc) & (AU / yr)                    \\
\noalign{\smallskip}
\tableline
\noalign{\smallskip}
J0543+2329  & B0540+23& 184 & -3.3 & 3.5 & 80\\
J0826+2637 & B0823+26& 197 & 32   & 0.4 & 40\\
J1136+1551 & B1133+16 &242 & 69   & 0.4 & 130\\
J1740+1311 & B1737+13 &37   & 22   & 4.7 & 140\\
J1932+1059 & B1929+10 & 47  & -3.9 & 0.3 & 40\\
J2018+2839 & B2016+28 & 68  & -4.0 & 1.0 & 7\\
\noalign{\smallskip}
\tableline
\end{tabular}
}
\end{center}
\end{table}

For each pulsar, we differenced the six pairs of absorption spectra corresponding
to the four observing sessions. We found few statistically significant optical depth 
variations.  The  rare cases of significant changes were mostly seen toward PSR 
B1929+10 and PSR B2016+28 (see Figs. 1 and 2).  The detections result from  variations 
having column densities of $\sim10^{18-19}$ cm$^{-2}$, with implied volume densities 
(for spherical clouds) of $\sim10^{4-5}$ cm$^{-3}$.  It is worth noting that the PSR B1929+10
fluctuations have a factor of 3-4 lower column densities than the ``canonical'' HI TSAS discussed
by \citet{h97}.

\begin{figure}[!ht]
\includegraphics[scale=0.6,trim=-1in 0 0 0 ]{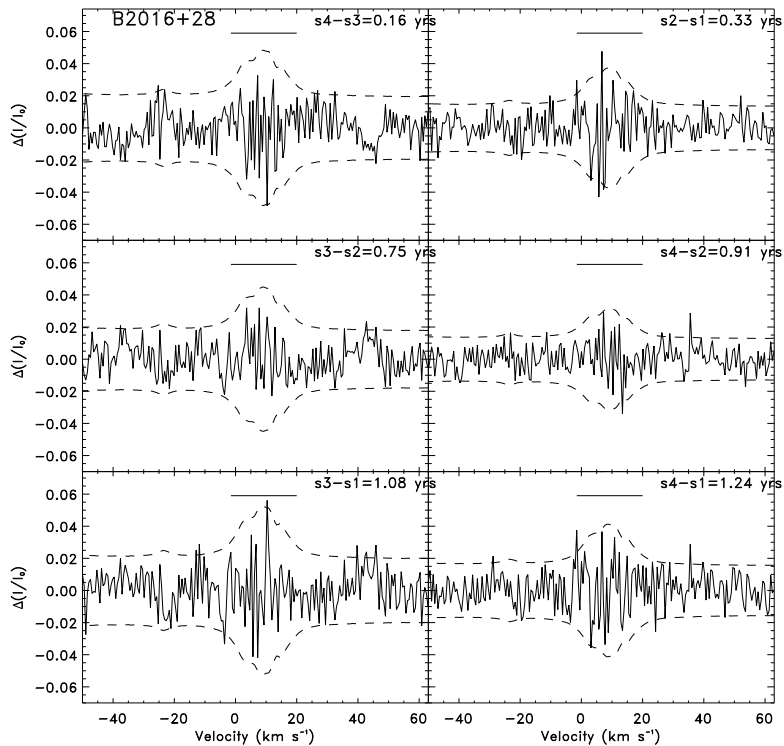}
\caption{Differences between pairs of absorption spectra for B2016+28.  See Fig. 1 for details.}
\end{figure}

We have also studied the {\it{integrated}} properties of the variations using equivalent width EW,
where EW=$\int\tau dv$.  Fig. 3 shows $|\Delta$EW$|$ versus EW for all our pulsars except B0540+23,
whose optical depth integrals become very uncertain near $\tau\sim\infty$ features.  PSR 1929+10
exhibits the only significant $(>3\sigma)$ nonzero values of $|\Delta$EW$|$.  Our results on PSRs B1737+13  and particularly B2016+28 [primarily upper limits with an occasional marginal 
$(<2\sigma)$ detection] are far lower than the detections of Frail, Weisberg, Cordes,  
\& Mathers (1994) [$|\Delta$EW$|\leq0.2 \ vs.\  |\Delta$EW$|\sim 0.5$ for B1737+13 and 
$|\Delta$EW$|\leq 0.3\ vs. \ |\Delta$EW$| \sim(1-5)$ for B2016+28].  We believe that at least 
some of the variations they saw were due to small calibration errors.  Our results do not rule out 
their finding of a correlation between $|\Delta$EW$|$ and EW, but the evidence for such a trend
appears marginal in our data.  

\begin{figure*}[!ht]
\includegraphics[scale=0.7,trim=0 0.6in 0 0]{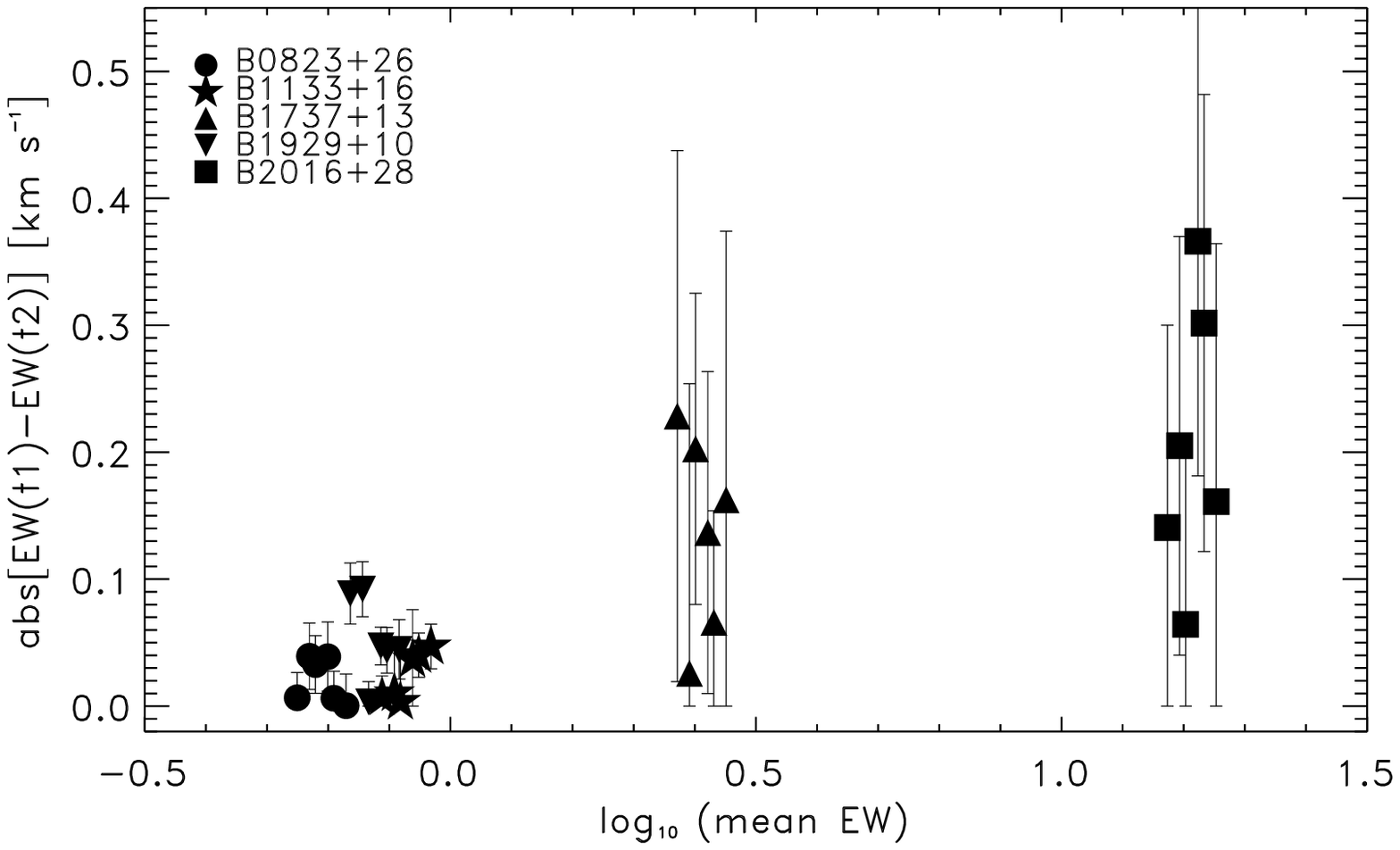}
\caption{Equivalent width changes $|\Delta$EW$|$ versus EW in the pulsar HI experiment 
of \citet{s07}, where equivalent width EW=$\int\tau dv$.}
\end{figure*}

\subsection{A Sketch of Recent Interferometric HI TSAS Experiments}

The state of the art of interferometric measurements of HI TSAS has also advanced significantly
since the first detection of \citet{d76}, especially with the advent of the VLBA.  The sources 3C 147 
and 3C 138 show the most prominent
fluctuations.   \citet{fg01}, who have done the most recent observations and analyses of 
3C 147, find $\Delta\tau\sim$ 0.3 over $\sim 25$ AU scales. The newest 
results on  3C 138  by \citet{b05} indicate variations of $\Delta\tau\sim0.5$ on 25 AU scales.  
The fluctuations  appear in approximately 10\% of their pixels on the source. Most modern 
interferometer results on other sources yield  smaller and
less frequent optical depth variations \citep{f98}.  See the contribution by Brogan et al
in this volume for a thorough review of these measurements.

\subsection{Discussion of HI TSAS Results}

The various pulsar and interferometric measurements reviewed above appear to lead to divergent
conclusions about the properties and pervasiveness of HI TSAS. Yet Occam's Razor suggests that
we search for the simplest explanation that can explain them all.

Among the pulsar measurements, it appears that \citet{f94}, who found that TSAS is rather 
pervasive,  provide the principal outliers.  We have noted above that there appear to be calibration
issues in these measurements which may have led to overestimates of the prevalence of 
optical depth variations.  In our subsequent analyses, we have replaced their results with our
newer measurements of the same pulsars \citep{s07}.

\citet{b05} note that the two sources exhibiting the most significant evidence for TSAS among HI 
interferometer results, 3C 138 and 3C 147, are also the only two with angular sizes significantly 
larger than the detected gas fluctuations.  They propose that the lower measured levels of 
variations in the other  cases results from  a selection
effect due to incomplete sampling of the relevant angular and spatial scales.  They also attempt
to model pulsar results by studying the statistics of  one-dimensional spatial cuts across the face
of 3C138.  They conclude that the relatively low level of recent pulsar TSAS detections can
also be explained by an incomplete sampling of angular scales due to the episodic nature of
the observations.  However, their model pulsar observations are all separated by the same 
angular scale, namely the characteristic scale of the HI fluctuations; whereas multiepoch
observations of a given pulsar actually sample a variety of scales.  The recent observations
of \citet{mET05}, published after the \citet{b05} analysis was complete, are particularly notable
in finding no fluctuations over a very wide and almost continuous range of scales.

In an attempt to make progress toward a comprehensive explanation of the various
observations, we have plotted HI TSAS detections and upper limits as a function of their
spatial scale in Fig. 4.  The expected level of optical depth variations as a function of scale 
 calculated from the \citet{d00}  power law fluctuation model  is also shown as a sloping line.
As these values are extrapolated from much larger observed scales, they could be significantly
in error but the slope is more secure.

\begin{figure*}[!h]
\includegraphics[scale=0.6,trim=-0.3in 0.5in 0 0]{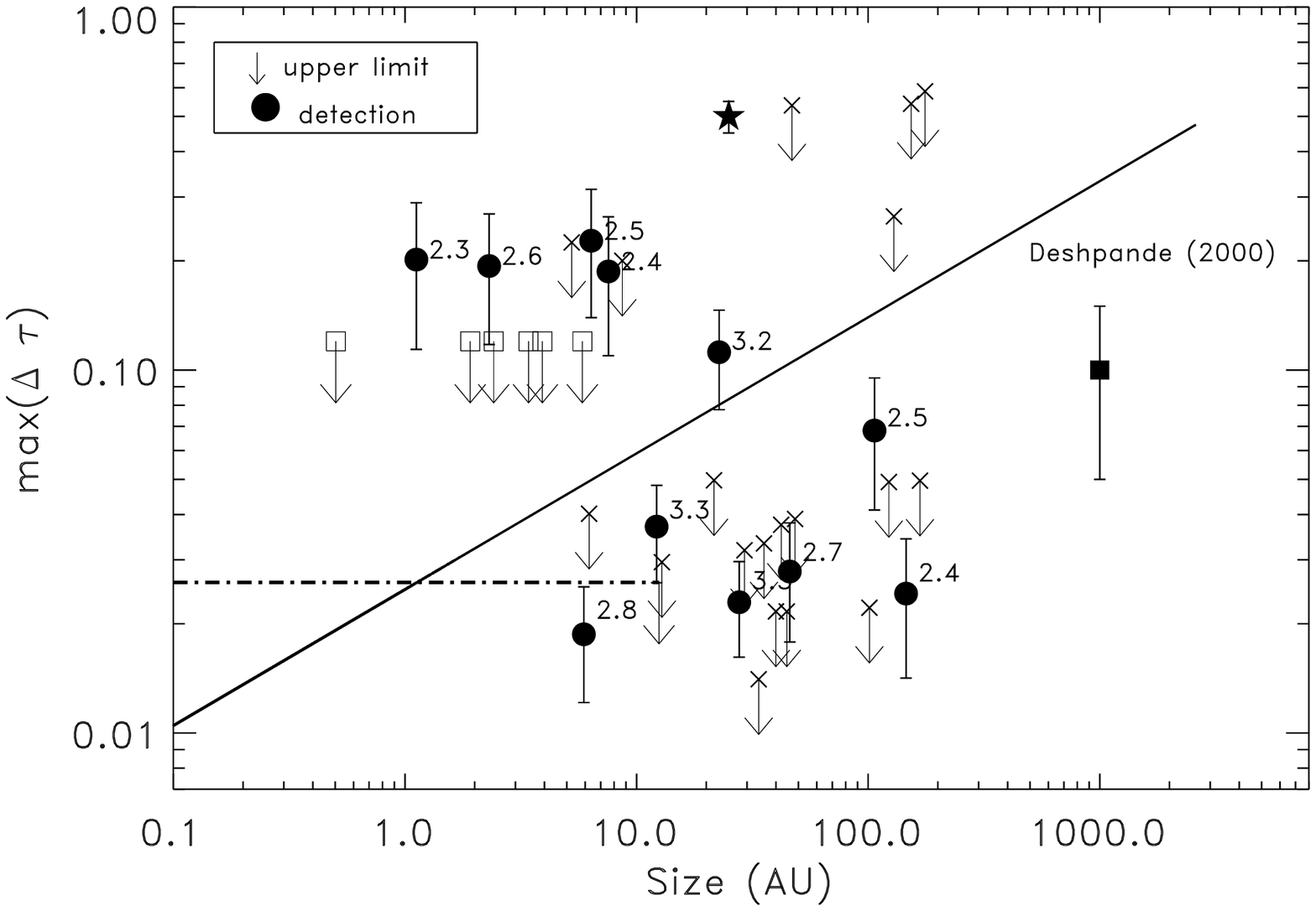}
\caption{HI optical depth fluctuations as a function of spatial scale. The  \citet{b05} 
interferometric 3C 138 
fluctuations (star), the \citet{j03} pulsar detection (filled square) and upper limits (open squares),
the  \citet{mET05} pulsar upper limits (dot-dashed horizontal line), and \citet{s07} pulsar detections
(filled circles) and upper limits (crosses) are shown, along with the \citet{d00} theoretical 
expectation  (sloping line). The depth of the fluctuation in units of $\sigma$ is placed next to each
detection.}
\end{figure*}

We note that the \citet{b05} interferometric detection is at a higher level than any other, which
 suggests that the
line of sight may indeed have a higher level of fluctuations than is typical.  Note however
that several of our pulsar detections  
at slightly smaller scales (mostly PSR 2016+28), though exhibiting somewhat 
lower optical depths, are consistent with the \citet{b05} detection  if the \citet{d00} power law
slope is correct.  Meanwhile numerous other pulsar detections and upper limits lie well below
the above mentioned ones; most especially our PSR B1929+10 results and 
the \citet{mET05} upper limits over a wide range of scales.
The figure also emphasizes that pulsar
TSAS experiments have made detections on many spatial scales, whereas the two most secure 
interferometric results both yield a single scale of $\sim25$ AU \citep{b05}. 
Hence we conclude that the observed variations in level and scale of HI TSAS are more 
than a selection effect, and that they are strongly dependent upon local conditions along a
particular line of sight.  

Recently \citet{bk05} and \citet{sh05} have detected a population of small ($10^{3-4}$ AU)
 HI clouds via very 
sensitive  absorption measurements of extragalactic background sources.  The optical 
depths and column densities range below 1 \% and $10^{18}$ cm$^{-2}$, respectively.  
It is not yet clear whether these features represent a distinct population or merely an extension 
of the pulsar-detected TSAS
toward larger spatial size and smaller optical depth.

We have examined the lines of sight toward our two pulsars with the most significant fluctuations.
Nothing particular was found in the direction of PSR B2016+28.  However, the line of sight toward
PSR B1929+10 appears to run {\it{along}} the wall of the Local Bubble for a significant stretch, 
according to the map of \citet{l03}. While the Local Bubble itself is probably  a relatively 
quiescent region, its boundary  is thought to be marked by enhanced turbulence \citep{pc92}.  We
currently regard this finding as potentially important but tentative, and we await further refinements 
in knowledge of the location of the Local Bubble boundary.

\section{Studies of  Molecular Tiny-Scale Structure}

 The tiny-scale structure of molecular gas in the ISM is an interesting area for investigation, both in its
 own right and in comparison with neutral atomic hydrogen (\S3).  Interstellar molecular spectral lines 
 usually have  much  smaller optical depths than HI, making them  harder to detect. Therefore most
 successful radio observations of small scale structure in molecular gas awaited relatively recent 
 advances  in sensitivity and angular resolution.  
 
 \citet{mmb93} and \citet{mm95} did a series of molecular absorption line experiments using
 extragalactic background sources.  They detected significant  time-variable 
 4.8 GHz H$_2$CO  absorption with $\Delta\tau=0.02-0.03$  in front of two compact sources, 
 which they  interpret as resulting from the superposition of several clumps along the line
 of sight whose size is  $\sim10$ AU or less, 
 having HI column densities of  $\sim10^{20}$ cm$^{-2}$ and volume densities of
 $\sim 10^6$ cm$^{-3}$.  
 
 Pulsars can be excellent probes of molecular gas for all the reasons given in \S2, providing
 only that the pulsar signal is sufficiently strong at the line frequency.  However, pulsars
 have rather steep spectra, so the only lines matching this criterion are the four 18 cm lines of
 OH.

 The first published pulsar OH ISM experiments were done by  \citet{s72} and \citet{g74}.  Both 
 of these  investigators  detected no absorption in the spectrum of PSR B0329+54; while Slysh 
 detected no absorption in several others.  His detection of OH absorption at 1667 MHz in PSR 
 B1749-28 was not confirmed by later observations.  

 Taking advantage of the high sensitivity of Arecibo, we and our colleagues did the first modern 
pulsar studies of interstellar OH \citep{s03b}.  Among the seven pulsars studied, only PSR 
B1849+00 clearly exhibits OH absorption.  This line of sight is remarkable in that it passes 
near the edge of  the SNR Kes 79 (G33.6+0.1).  
 Fig. 5 displays our OH spectra  gathered in this direction.  The absorption lines in the  
 pulsar-off spectra (bottom row) result from
 SNR continuum radiation being absorbed by foreground gas.  Note that these pulsar-off
 absorption lines exhibit much smaller optical depths  than do the pulsar OH spectra (top row) 
 at $v\sim 100$ km s$^{-1}$ ($\tau\sim 0.05 \ vs.\  0.4$ at 1665 MHz and $\tau\sim0.02\ vs.\ 0.9$ at
1667 MHz).  These smaller pulsar-off optical depths are much 
more representative of typical OH absorption optical depths
in the plane (e.g., Dickey et al 1981), so that our measured pulsar OH optical
depths are anomalously high.  \citet{s03b} discuss several possible explanations for the high optical
depths in the pulsar OH interstellar spectra.  The most likely is that the tiny column through the ISM 
sampled by the pulsar signal has encountered a small ($<15\arcsec$), high density 
($> 10^5$ cm$^{-3}$) molecular cloudlet.  Meanwhile, the 
pulsar-off spectrum, in sampling absorption across the full telescope beam, represents a solid-angle 
average across a clumpy medium consisting of relatively high optical depth, small  molecular cloudlets, 
 embedded in a lower density medium. 
 
 \begin{figure*}[h!]
\includegraphics[height=3in,width=5in,trim=-0.5in -0.5in 0 0]{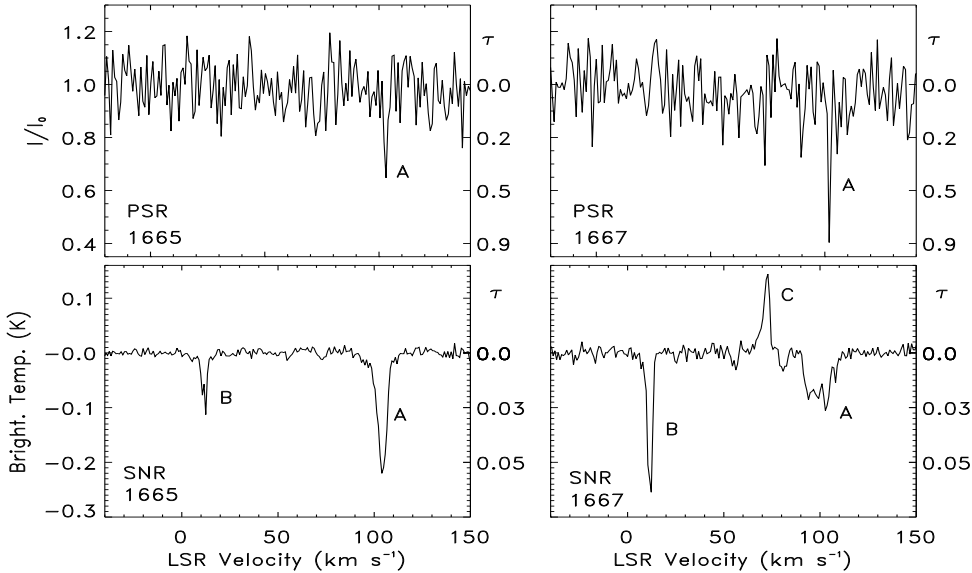}
\caption{Arecibo OH spectra toward PSR B1849+00, after \citet{s03b}. The top row shows the
pulsar ISM spectra at 1665 and 1667 MHz. The bottom row, taken when the pulsar is off,
shows emission and absorption throughout the telescope beam, which is filled by
the SNR Kes 79 (G33.6+0.1). Compare pulsar ISM optical depths (top) with pulsar-off, SNR
optical depths (bottom). }
\end{figure*}

 \begin{figure*}[!ht]
\includegraphics[height=6in,width=5.5in,trim=0 0 0 0  ]{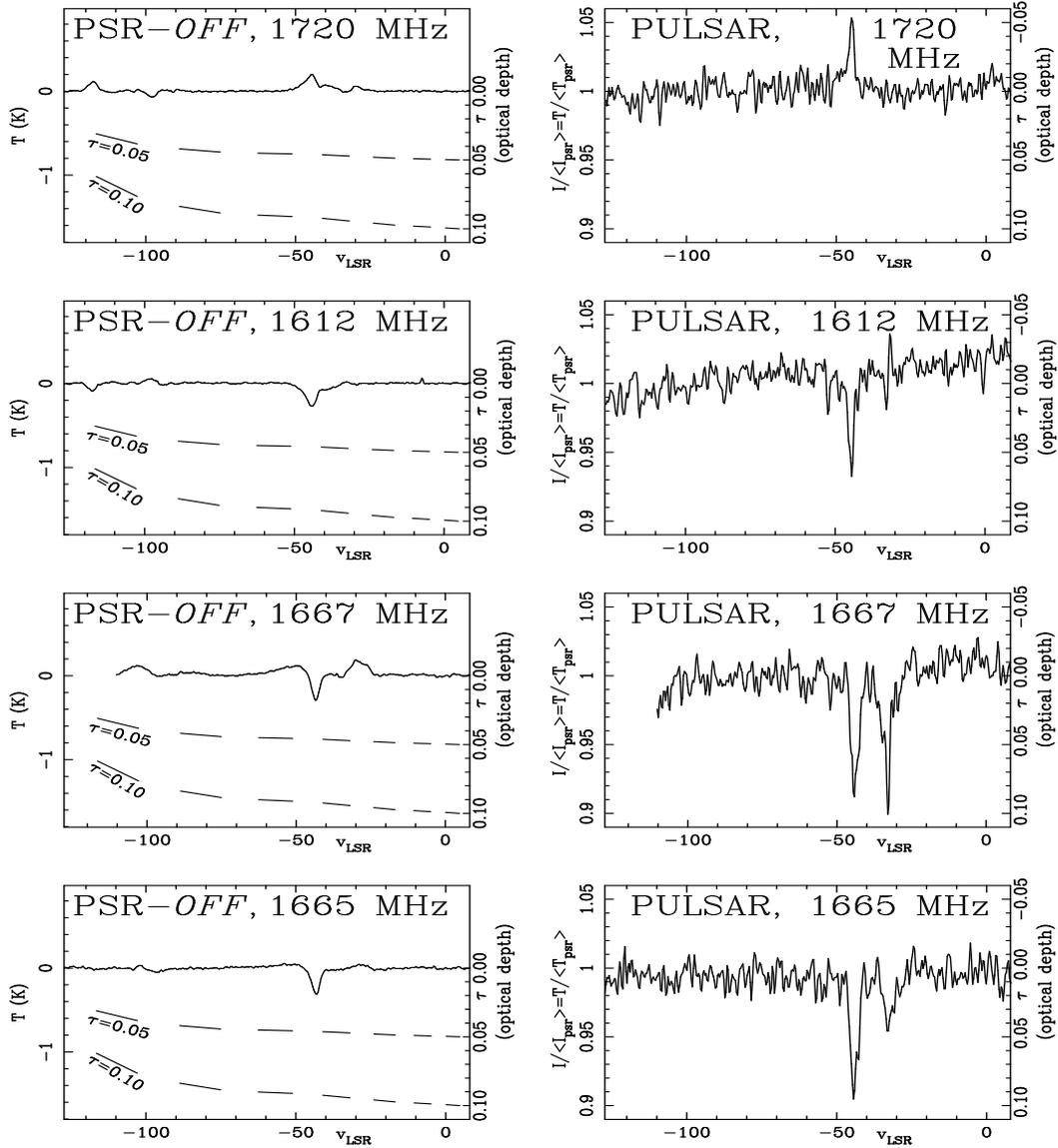}
\caption{Pulsar-off (left) and Pulsar ISM (right)  spectra at the four 18 cm OH lines
toward PSR B1641-45, from Parkes observations by \citet{w05}. All spectra are plotted
with the same optical depth scale.  Compare pulsar-off and pulsar ISM optical depths, and 
also note the pulsed interstellar OH maser stimulated by the pulsar at 1720 MHz.} 
\end{figure*}

\citet{w05} continued the search for additional pulsar OH interstellar lines with the Parkes
telescope.  Of eighteen pulsars searched, only B1641-45 exhibited detectable pulsar ISM
spectra (see Fig. 6).  Note that the pulsar ISM spectra (right column) exhibit significantly larger
optical depths than do the corresponding pulsar-off spectra (left column), continuing the trend 
discussed above in PSR B1849+00. So it again appears that the pulsar-Earth pencil beam 
has intercepted  dense cloudlets in the pulsar ISM spectra, whose properties suffer angular 
dilution in the  pulsar-off spectra.  Note also that stimulated emission is present in the 1720 
MHz pulsar ISM spectrum.  This is the first detection of a pulsed interstellar maser. Since
the stimulated emission appears in the spectrum of the pulsar alone, it represents  a direct
demonstration that the emission is stimulated by pulsar photons.

\citet{m05} has also recently searched for OH lines in the spectra of five pulsars with the GBT.
He detected absorption toward PSR B1718-35.

\subsection{Discussion of Molecular Results}

Theory suggests that molecular clouds are a dynamic, turbulent environment \citep{mk04,bET06}.
\citet{bh02} performed a principal component analysis of CO maps,  finding a  slope of 2.17 for the energy
spectrum, which they interpreted as evidence for continuous energy injection at small scales.  Thus it is not surprising
that both the formaldehyde results and our groups' pulsar OH work indicate the presence of tiny-scale 
structure in molecular gas.  Our conclusions result from our detection of larger optical depths in
the pulsar ISM spectra than in the pulsar-off spectra.  Multiepoch experiments are now in progress
to search for time variability in the these lines, much like the pulsar HI experiments of \S3.1 and 
3.2.  Detection of time variability will provide powerful additional 
support for the existence of TSAS in OH.

\acknowledgements JMW and SS gratefully acknowledge  support from NSF Grants 
AST-0406832 and AST-0406987, respectively. Arecibo Observatory is part of the National Astronomy
and Ionosphere Center, operated by Cornell University under a cooperative agreement with
the NSF.  The Green Bank Telescope of the National Radio Astronomy Observatory is a facility of the 
NSF operated under cooperative agreement by Associated Universities, Inc.  The  Parkes telescope
is part of the Australia Telescope which is funded by the Commonwealth of 
Australia for operation as a National Facility managed by CSIRO.

\end{document}